# Environmentally-Extended Input-Output analyses efficiently sketch large-scale environmental transition plans: illustration by Canada's road industry

Anne de Bortoli[1,2,3,4]*, Maxime Agez[1]

**1** CIRAIG, École Polytechnique de Montréal, P.O. Box 6079, Montréal, Québec, H3C 3A7, Canada

**2** Americas Technical Center, Eurovia Canada Inc., 3705 Place Java #210, Brossard, QC J4Y 0E4, Canada

**3** Technical Direction, Eurovia Management, 2 rue Thierry Sabine, 33700 Mérignac, France

**4** LVMT, Ecole des Ponts ParisTech, Université Gustave Eiffel, 6-8 Avenue Blaise Pascal, 77420 Champs-sur-Marne, France

**\*** Corresponding author; e-mail: anne.debortoli@polymtl.ca

ABSTRACT

Industries struggle to build robust environmental transition plans as they lack the tools to quantify their ecological responsibility over their value chain. Companies mostly turn to sole greenhouse gas (GHG) emissions reporting or time-intensive Life Cycle Assessment (LCA), while Environmentally-Extended Input-Output (EEIO) analysis is more efficient on a wider scale. We illustrate EEIO analysis' usefulness to sketch transition plans on the example of Canada's road industry: estimation of national environmental contributions, most important environmental issues, main potential transition levers of the sector, and metrics prioritization for green purchase plans. To do so, openIO-Canada, a new Canadian EEIO database, coupled with IMPACT World+ v1.30-1.48 characterization method, provides a multicriteria environmental diagnosis of Canada's economy. The construction sector carries the second-highest environmental impacts of Canada (8-31% depending on the indicator) after the manufacturing industry (20-54%). The road industry generates a limited impact (0.5-1.8%), and emits 1.0% of Canadians' GHGs, mainly due to asphalt mix materials (28%), bridges and engineering structures materials (24%), and direct emissions (17%). The industry must reduce the environmental burden from material purchases - mainly concrete and asphalt products - through green purchase plans and eco-design and invest in new machinery powered with cleaner energies such as low-carbon electricity or bioenergies. EEIO analysis also captures impacts often neglected in process-based pavement LCAs - amortization of capital goods, staff consumptions, and services – and shows some substantial impacts advocating for enlarging system boundaries in standard LCA. Yet, pavement construction and maintenance only explain 5% of the life cycle carbon footprint of Canada's road network, against 95% for the roads' usage (72% from vehicle tailpipes releases, 23% for manufacturing vehicles). Thereby, a carbon-neutral pathway for the road industry must first focus on reducing vehicle consumption and wear through better design and maintenance of roads. Finally, EEIO databases must be developed further as a powerful tool to fight planet degradation, and openIO-Canada must be specifically expanded and refined to allow for more robust and larger multicriteria assessments.

# 1 Introduction and background

The rise of environmental threats calls for a drastically hurried industrial environmental transition. Building an efficient action plan at the industry level requires a deep multicriteria understanding of environmental life-cycle impacts to determine the most effective levers. But the tools and data to reach sufficient knowledge seem unsatisfying for the industry.

First, the compilation of sectoral national greenhouse gas (GHG) inventories can be an interesting source of data to understand the carbon structure of an economy and its industry. However, GHG emissions are only one environmental dimension. An improvement towards reducing these emissions could mask worse performance in other environmental categories, as a consequence or independently of the taken measures. Moreover, these inventories are not suitable for industries as they only consider the emissions of production taking place on the territory - an approach known as "production-based" or "territorial" (Caro et al., 2014). Territorial approaches help decreasing impacts on a given territory but tend to push governments to shift the emissions outside the country by relocating polluting practices (Peters and Hertwich, 2008) and importing the most carbon-intensive products, which can lead to increases in the global emission levels of GHGs and degrade the national economy (Rieber and Tran, 2008). Therefore, these analyses must be complemented with supply-chain-wide analyses accounting for foreign impacts. Such analyses, called national "consumption-based" environmental assessments – i.e. accounting for the impact from cradle-to-consumer, disregarding the production location - remain limited (Harris et al., 2020), even though they seem particularly suitable to orientate sustainable purchases and productions accounting for the entire supply chain.

Given the methodological choices made for national accounting, industrial sectors and companies face three important challenges to determine their impact on the environment. First, because national inventories follow the territorial approach, companies do not have easy access to data on their indirect emissions - called "scope 3" assessment according to the GHG protocol (WBCSD and WRI, 2004) - as most of these indirect emissions occur beyond the borders of one single territory. Yet, the assessment of the so-called scope 3 is crucial to implementing comprehensive impact reduction procedures, such as green purchasing policies and environmental research and development (R&D) plans. For instance, in pavement construction, bitumen constitutes a major source of environmental impacts (de Bortoli, 2020) and falls within scope 3 according to the GHG Protocol definitions. Second, the governmental services responsible for national GHG inventories often perform aggregations of the emissions by broad categories before releasing environmental information (Government of Canada, 2021a). This further hinders the potential use of this data by companies. Indeed, data is too aggregated to allow specific industrial sectors to access a representative footprint of their activities. As an example, to our knowledge, the road construction sector is not considered as such in national inventories, and even less the more specific sectors of materials it produces or consumes (i.e., aggregates, cement, binders, etc.). Third, national inventories only provide a footprint, most of the time solely focusing on GHG emissions. It does not describe intersectoral links which

https://doi.org/10.1016/j.jclepro.2023.136039, accepted version, Journal of Cleaner Productionwould allow companies to identify the main sources of impact in their supply chain and lack a multicriteria comprehensiveness.

While national inventories are mostly focused on GHG emissions, as well as water and energy use, some countries also register other pollutant emissions by economic sector, such as the US EPA through its Chemical Data Reporting (US EPA, 2011), or the French CITEPA (CITEPA, 2021). However, these environmental matrices only report flows, i.e., masses of substances emitted which need further analyses before being usable by companies, to determine what the environmental impacts of these pollutants is. These flows thus need to be transformed from emissions to impacts, through characterization factors that model fate, exposure, and effect of these pollutants in the natural environment.

Two environmental quantification methods overcome the limitations previously raised: Life Cycle Assessment (LCA) and Environmentally Extended Input-Output (EEIO) analysis. Both carry out multi-criteria appraisals – e.g., focus on several environmental impact categories - adapted to industry transition plans, that is, from cradle-to-gate, i.e., from the extraction of materials to the products leaving the factory. However, using LCA to analyze the environmental impacts of a sector at the national level requires an enormous amount of data, thus triggering the intensive time consumption of data collection, which is one of the main obstacles to the large-scale application of LCA. Moreover, the lack of digitization in certain sectors such as the construction sector can hinder access to the required data. On the other hand, EEIO analysis allows for such large-scale analyses, and databases to perform them constantly emerge (Huo et al., 2022; Agez, 2021; Stadler et al., 2018; Yang et al., 2017; Andrew and Peters, 2013; Lenzen et al., 2013, 2022). These two methods fulfill different roles. EEIO enables large scale assessments thanks to its complete coverage of the economy, while LCA mainly focuses on supply chain assessments through its much more supply-chain specific data. Hybrid LCA - completing LCA with EEIO analysis - is also an interesting tool to fill the gaps of what can be difficultly model through a simple process-based LCA due to a lack of data (LCIs or foreground data).

EEIO analysis is already used within academia to quantify the environmental contributors to national economic sectors with a consumption-based approach. However, EEIO applications either focus on the consumption of final demand sectors (e.g., households, government) (Castellani et al., 2019; Cellura et al., 2011), or the impact of a whole economy typically without going into intersectoral details (Dawkins et al., 2019; Lenzen, 2007; Moran et al., 2018; Xia et al., 2022). Regarding applications to specific industrial sectors, the literature has so far mainly focused on food (Reutter et al., 2017) or tourism (Sun et al., 2020). But to our knowledge, there is no application in the road industry, only one EEIO analysis focusing on the GHG emissions from the entire construction sector in Ireland (Acquaye and Duffy, 2010). EEIO analysis is therefore a powerful decision support tool but is often overlooked outside of academia. It could be used more widely to allow industries to build environmental transition plans by identifying major environmental contributors, selecting relevant sets of environmental indicators to monitor, prioritizing R&D topics, building green purchasing policies, and identifying major improvement levers for production sites.

The overall objective of this article is to illustrate the possible uses of EEIO analysis to build environmental transition plans for companies and industries on a wide scale such as a nation. This illustration will be based on the example of the road industry in Canada. We will first present the EEIO analysis method, and the database used for the Canadian context (section 2), before detailing the calculations that will be carried out (section 3). Then, the results (section 4) include a cross-sectoral vision of Canada's multicriteria environmental impacts (12 midpoint indicators, 2 endpoint indicators) and the contribution of the construction sector, an intra-sector environmental picture of the road industry highlighting direct transition levers, and finally an overview of its impact by scope to give a perspective on these levers. In the last section, the results will be discussed to lead to some recommendations to decision-makers before detailing the benefits and limits of using EEIO analysis instead of LCA on large scale systems (section 5).

## 2   Method and material

### 2.1   The EEIO analysis method

EEIO analysis, also shorten as "EEIO", is a method that links environmental consumptions or releases to monetary transactions (Miller and Blair, 2009). Data (both economic and environmental) are obtained and compiled by national statistics agencies, often through surveys sent to companies. Academics and government agencies then integrate this data to form tables that can be readily used to estimate emissions and impacts on the environment. This quantification accounts for emissions in a cradle-to-gate fashion. Hence, it includes scopes 1 and 2, as well as upstream scope 3. Although generally focused on the climate change issue, it can cover a large range of pollutants (i.e., 35 pollutants in the EXIOBASE EEIO database (Stadler et al., 2018)), mineral resources and land use and thus can be used to study other environmental issues, such as acidification or eutrophication.

Impacts on the environment linked to a given final demand (e.g., households) are determined using the following equation:

$$q = C \cdot B \cdot (I - A)^{-1} \cdot y \qquad (3)$$

Where vector $q$ presents the different potential impacts on the environment, $C$ is a matrix regrouping characterization factors translating emissions to potential impacts on the environment, matrix $B$ regroups the environmental extensions linking estimated consumptions and/or emissions for 1\$ of each category of product, $I$ is the identity matrix, $A$ is the technology matrix describing the normalized monetary exchange between sectors of the economy, and vector *y* being the final demand.

## 2.2 Sources of Canadian data

OpenIO-Canada is the EEIO database used in this article (Agez, 2021). It is the first open-source EEIO database developed for Canada. The database represents the whole Canadian economy while providing details at the provincial level, for years from 2014 to 2017. Hence, the model includes economic exchanges between the different provinces as well as specific consumptions and polluting emissions for each economic sector of each province. The economic data of openIO-Canada comes from the Supply and Use tables provided by Statistics Canada (StatCan) (Statistics Canada, n.d.), using a classification breaking down each province's economy in 492 commodities. The database is a single region input-output database and as such models international trade with local data, thus applying a domestic technology assumption. It is therefore assumed that production abroad has the same intensity as in Canada. GHG emissions and water use data are derived from the physical flow accounts of StatCan, while the 323 other types of pollutants come from the National Pollutant Release Inventory (NPRI) (Government of Canada, 2021b). Physical flow accounts are provided with a lower level of detail than the economic accounts. (e.g. 240 vs 492 sectors for GHG emissions and water flows). An economic allocation is thus performed to distribute the environmental flows between the different products of the economic sectors. Moreover, we attribute to imported products the same environmental flows as similar goods produced in the importing region, as is typically done in many national EEIO tables (Eurostat : Statistisches Amt der Europäischen Gemeinschaften, 2008).

# 3   Calculations

## 3.1   Overview

Figure 1 gives an overview of the calculation process used to understand the sources of the environmental impacts of Canada and its road sector in 2017, based on the latest data made available by the Canadian Government. A consumption-based approach is adopted throughout this paper, i.e., impacts stemming from international exports are excluded while international imports, local production, and consumption are included.

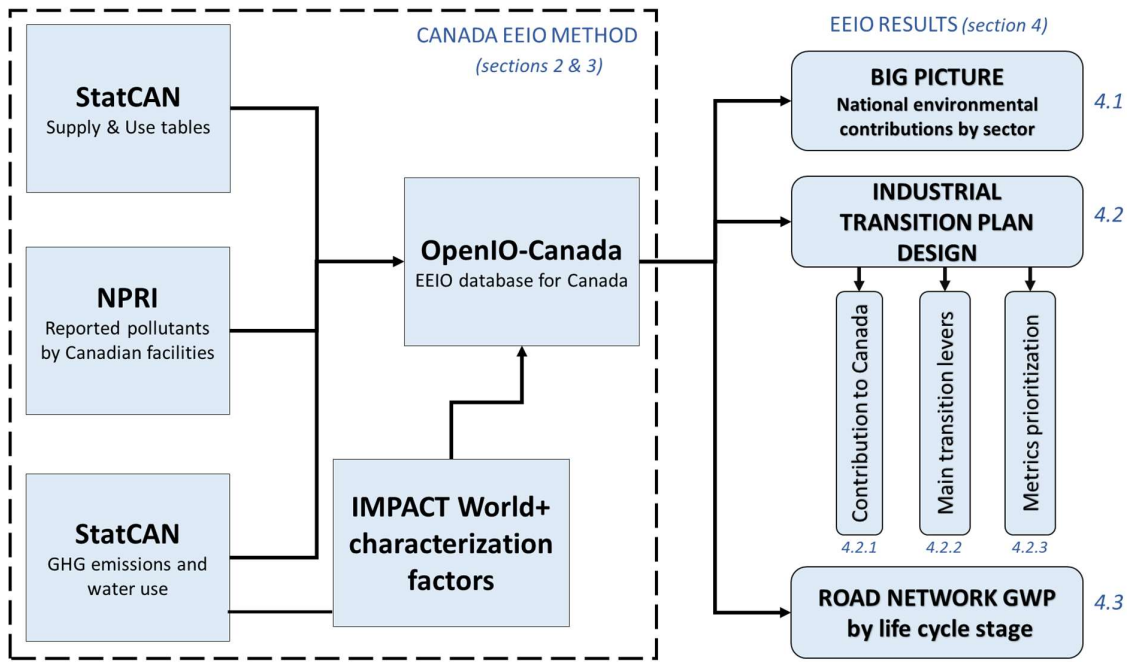

**Figure 1 Illustration of the calculation method framework and presented results**

## 3.2   Characterization method and indicators

The life cycle impact assessment method IMPACT World+ (IW+) v1.30-1.48 was selected as it was deemed the most scientifically up-to-date and relevant method for Canada (Bulle et al., 2019). Indeed, the TRACI method recommended as a by-default method in many North American LCA standards, such as the ISO 21930 standard dedicated to construction LCA (International Organization for Standardization, 2017), is outdated (e.g. reduced set of ten midpoint indicators offering a less holistic environmental coverage, lack of endpoint indicators). Moreover, IW+ is one of the most advanced LCIA method, calculating endpoint scores from elementary flows based on causal chains, rather than applying factors to midpoint indicators as performed in the other LCIA methods. Both midpoint and endpoint indicators were calculated. At the midpoint level, indicators calculated were short-term climate change (also called Global Warming Potential (GWP)), freshwater acidification, eutrophication and ecotoxicity, marine eutrophication, ozone layer depletion, particulate matter (PM) formation, photochemical oxidant formation, terrestrial acidification, carcinogenic and non-carcinogenic human toxicity resp. "cancer" and "non-cancer"). As for the endpoint indicators, the damage to ecosystem quality and human health are considered. Both the long- and short-term effects of climate change on these damages are considered in IW+. Due to the lack of available national data, some IW+ indicators were not considered, such as ionizing radiations, resources (mineral and fossil), land transformation and occupation, as well as water scarcity. Nevertheless, an indicator of water use was calculated, using the data reported by Statistics Canada (n.d.). Water use only accounts for withdrawn water and not released water. It is therefore different from water consumption indicators, which consist of the difference between withdrawn and released water.

### 3.3 *Aggregation of sectors by category*

This study relied on the "Detailed level" of openIO-Canada which describes the Canadian economy in 492 sectors. An additional category for direct emissions from final demands (e.g., coming from the use of personal vehicles by households) was also included. However, a 492-sectors classification being too much for a meaningful interpretation, contributions were aggregated. This aggregation was deemed necessary to make the results readable and usable for the orientation of road industry environmental transition plans, and it was made based on industrial road expertise and confirmations of some categories' meaning by the Statistical Information Service of Statistics Canada. The aggregation work was incremental, and the different aggregations tested are reported in the SM. The last two levels of aggregation are presented in Table 1 to explain better the results in Figure 4 and Figure 5. Disaggregated results (i.e., with the 492 sectors detail) are also available in SM.

**Table 1 Detail of the last two levels of agregation – levels 4 and 5**

| *Aggregation level #5* | *Aggregation Level #4* |
|---|---|
| Direct emissions | - |
| Bridges & tunnels | Concrete, cement, steel, aluminum, other metals |
| Asphalt mix materials | Asphalt products, crude oil, asphalt binders, aggregates, other minerals, chemicals |
| Energies | Electricity, natural gas, other fuels |
| Freight | Air, rail, road, water, and other freight services and supports |
| Infrastructure and capital goods | Other metals, machinery, plastic and rubber, buildings and infrastructure |
| Services | Administrative services, other services, upstream sales |
| Staff consumptions | Other goods, paper and paperboard, staff transportation |

# 4 Results and interpretation

## 4.1 Canada's environmental impacts

### 4.1.1 Environmental footprint contributions

Our analysis first focused on the diagnosis of the Canadian economy as a whole to get an overview of the environmental impact contributors in Canada and the environmental footprint of Canada and Canadians on a consumption-based approach. In 2017, Canada's consumption was responsible for the release of 692 million tons of $CO_2$eq (Mt$CO_2$eq). Considering a population of 38.4 million inhabitants (Statistics Canada, 2021), the average Canadian emits around 18.0 t$CO_2$eq-GWP100 per year. This is consistent with the values published by Friedlingstein et al. who reported a 15.7 tCO$_2$eq/capita.year carbon footprint in Canada in 2017 (Global Carbon Project, 2020) and EXIOBASE v3.8.2 which leads to 18.7 tCO$_2$eq/capita.year (own calculations with IW+ 1.30-1.48).

Figure 2 shows the contribution of 12 aggregated sectors of the Canadian economy – mainly following the classification of the economy by Statistics Canada (Statistics Canada, n.d.) - and 11 impact categories of the IW+ LCIA method, as well as a "Water use" indicator, on a consumption-based approach, encompassing the entire supply chain.

It shows that the manufacturing sector is by far the most important contributor to the environmental impact of Canadians' consumption on all midpoint impact categories. It goes from a minimum of 20% of the total water use to 54% of the impact on freshwater ecotoxicity, and accounts for a quarter of the impact of Canada on climate change. The construction sector is then globally the second most important contributor with contributions higher than 10% of the total (excluding on water use) and reaching up to 31% of the total on carcinogenic human toxicity. 12% of Canadians' impact on climate change is due to the construction sector, 41% of it being due to residential buildings. Services other than public administration and Non-Governmental Organizations (NGOs) are then ranked the third biggest contributor to the midpoint impacts, with a top contribution to water eutrophication (22%). Trade (i.e. wholesale, retail, warehousing activities, and their margins and commissions), as well as public administration and NGOs services, are two sectors that are visible contributors, explaining each 5 to 10% of the Canadians' impacts. Direct emissions related to Canadian consumption show negligible contributions apart from the climate change impact and the water use where it carries respectively 23 et 9% of the total impact. Direct GHG emissions of the Canadians' consumption represent 3.81tCO2e/capita. As the direct emissions of Canadian households solely come from GHGs emissions and water consumption – as the NPRI data do not cover households -, it is within expectations that it only contributes to these two impact categories.

Utilities present limited contributions (<5%) excluding on freshwater acidification (13%), marine eutrophication (11%), PM formation (9%), terrestrial acidification (12%) and water use (9%). Finally, agriculture, education, health, resources, and transportation sectors have low contributions, generally from 0 to 5%. Let's note that "Agriculture" accounts for agriculture, forestry, fishing, and hunting activities, and "Resources" encompass oil, gas, and mining. Transportation is a low contributor, including on GWP, because it only accounts for freight and public transportation directly bought by households. Indeed, household private transportation is encompassed within direct emissions. Besides, freight is not highly consumed by households and public transportation is rather low-emission. Finally, all other forms of transportation are included in the sector that buys them. For instance, impacts from freight services purchased by the construction sector are included in the environmental impact of this sector.

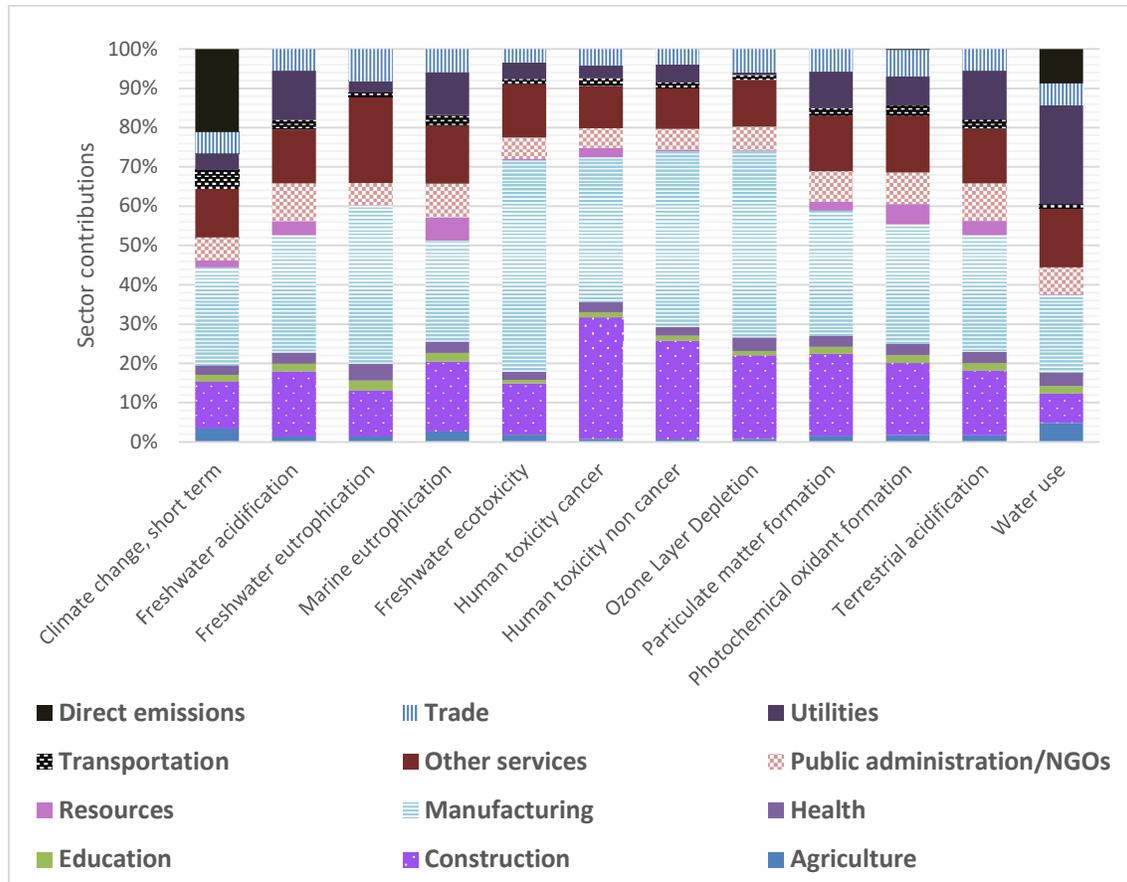

**Figure 2** Contributions of industrial sectors to the different IW+ environmental impacts of Canada at the midpoint level, consumption-based approach

*4.1.2   Most important environmental issues*

Based on the IW+ methodology, climate change – cumulating long- and short-term effects – is by far the most important midpoint contributor to the two damages: it brings respectively 95 and 82% of the human health and ecosystem quality damage (Figure 3). The rest of the human health damage is mainly brought by PM formation due to electricity consumption, while the ecosystem quality is otherwise mostly damaged by marine acidification, generated at 98% by $CO_2$ solubilization.

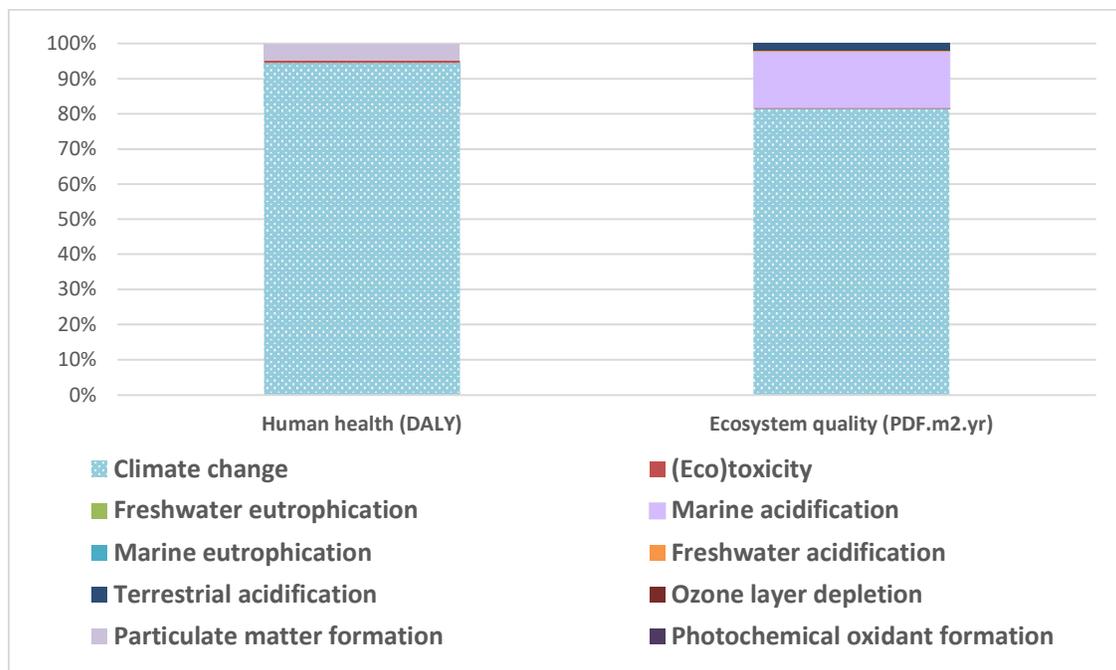

**Figure 3 Contributions of the different IW+ environmental impacts of Canada at the endpoint level, consumption-based approach**

## 4.2 Environmental transition plan for the road industry in Canada

In this section, we want to answer three main questions. First, what is the contribution of the road industry to the national environmental impacts of Canadians (section 4.2.1)? Second, what are the main potential levers to the environmental transition plan of the road industry (section 4.2.2)? And third, how to reduce the multicriteria dimension of green purchase strategies for decision-makers by prioritizing metrics (4.2.3)?

### 4.2.1 National burden from the road industry

Table 2 shows the contribution of the road sector to the total impact of Canada, as well as the contribution of the entire construction sector. It presents contributions within 0.54% to water use to 1.80% to particulate matter formation. In the climate change impact category, it accounts for 1.02%. It is also responsible respectively for 1.07 and 1.10% of the damage to ecosystems and human health. Overall, the contribution of the road sector to the national environmental impacts of Canada appears to be rather minor. Nevertheless, environmental preservation concerns now all sectors, whatever how important their impacts are, especially under the major risks generated by climate changes.

**Table 2 Contribution of the road and construction sectors to the total impact of Canada. The contribution of the road sector is included in the contribution of the entire construction sector.**

| INDICATOR | CONTRIBUTIONS PER SECTOR | |
|---|---|---|
| | ROAD SECTOR | ENTIRE CONSTRUCTION SECTOR |
| ECOSYSTEM QUALITY | 1.07% | 12.13% |
| HUMAN HEALTH | 1.10% | 11.82% |
| CLIMATE CHANGE, SHORT TERM | 1.02% | 16.56% |
| FRESHWATER ACIDIFICATION | 1.52% | 11.67% |
| FRESHWATER EUTROPHICATION | 0.68% | 17.82% |
| MARINE EUTROPHICATION | 1.64% | 13.10% |
| FRESHWATER ECOTOXICITY | 0.68% | 30.88% |
| HUMAN TOXICITY, CANCER | 1.28% | 25.11% |
| HUMAN TOXICITY, NON-CANCER | 1.67% | 21.17% |
| OZONE LAYER DEPLETION | 0.78% | 20.82% |
| PARTICULATE MATTER FORMATION | 1.80% | 18.34% |
| PHOTOCHEMICAL OXIDANT FORMATION | 1.43% | 16.52% |
| TERRESTRIAL ACIDIFICATION | 1.51% | 12.13% |
| WATER USE | 0.54% | 7.52% |

*4.2.2   Main potential levers for the transition*

**1.      Midpoint contribution analysis**

Figure 4 offers a first overview of the drivers of the environmental burdens generated by the road industry. We created categories based on our road industry expertise to aggregate the 492 sectors of openIO-Canada. The results illustrate that direct emissions are a low contributor to the cradle-to-gate impacts of the road industry, except for climate change where they represent 17% of the total impact. These low contributions might be due to a low number of road industry facilities reporting to NPRI.

The results also highlight that the category "bridges & tunnels" – a category including purchases of concrete, cement, steel, and aluminum – weighs heavily (from 11 to 60%) with regards to the majority of the environmental impacts of the road industry. Asphalt mixture materials purchases - in which we included the purchases in asphalt products, asphalt binders, aggregates, and other minerals as well as chemicals – highly affect 2/3 of the impact categories, particularly climate change (25%), freshwater acidification (29%), marine eutrophication (24%), ozone layer depletion (34%), PM formation (40%), photochemical oxidant formation (28%), and terrestrial acidification (28%). Energy purchases – including electricity, natural gas, and other fuels – account for 2% to 11% of the impacts (resp. for freshwater eutrophication and smog). On climate change, fuel purchase accounts for 9% of the impact and direct emissions for 17%, meaning a upstream scope 3 accounting for 32% of the climate change impact from energies, a higher contribution than the 10 to 15% usually highlighted in European databases (e.g. ADEME, 2019), probably due to the major supply from Canadian oil sands, known for their higher extraction and transformation impact (Charpentier et al., 2011; Chen et al., 2019; Masnadi et al., 2018). Infrastructure and capital goods – covering the maintenance/repairs of machinery, buildings, and infrastructure, as well as other metals than steel, plastic, and rubber – mainly have a significant impact (>10%) on freshwater ecotoxicity, human cancer, and non-cancer toxicities, ozone layer depletion and water use where they represent respectively 22, 15, 16, 18 and 15% of the contributions. Services – encompassing intragroup and external services as well as upstream sales –, have a prevalent role on the impact of freshwater eutrophication where they represent 56% of the impact. On other environmental indicators they contribute by 7 to 23%, thus being one the major contributors across all environmental indicators. Staff consumption – combining paper and paperboard purchases, other goods purchases, and staff transportation only has a significant impact (>10%) on freshwater eutrophication, freshwater ecotoxicity, human cancer and ozone layer depletion. Finally, freight services have negligible impacts except on climate change where they bring 5% of the emissions.

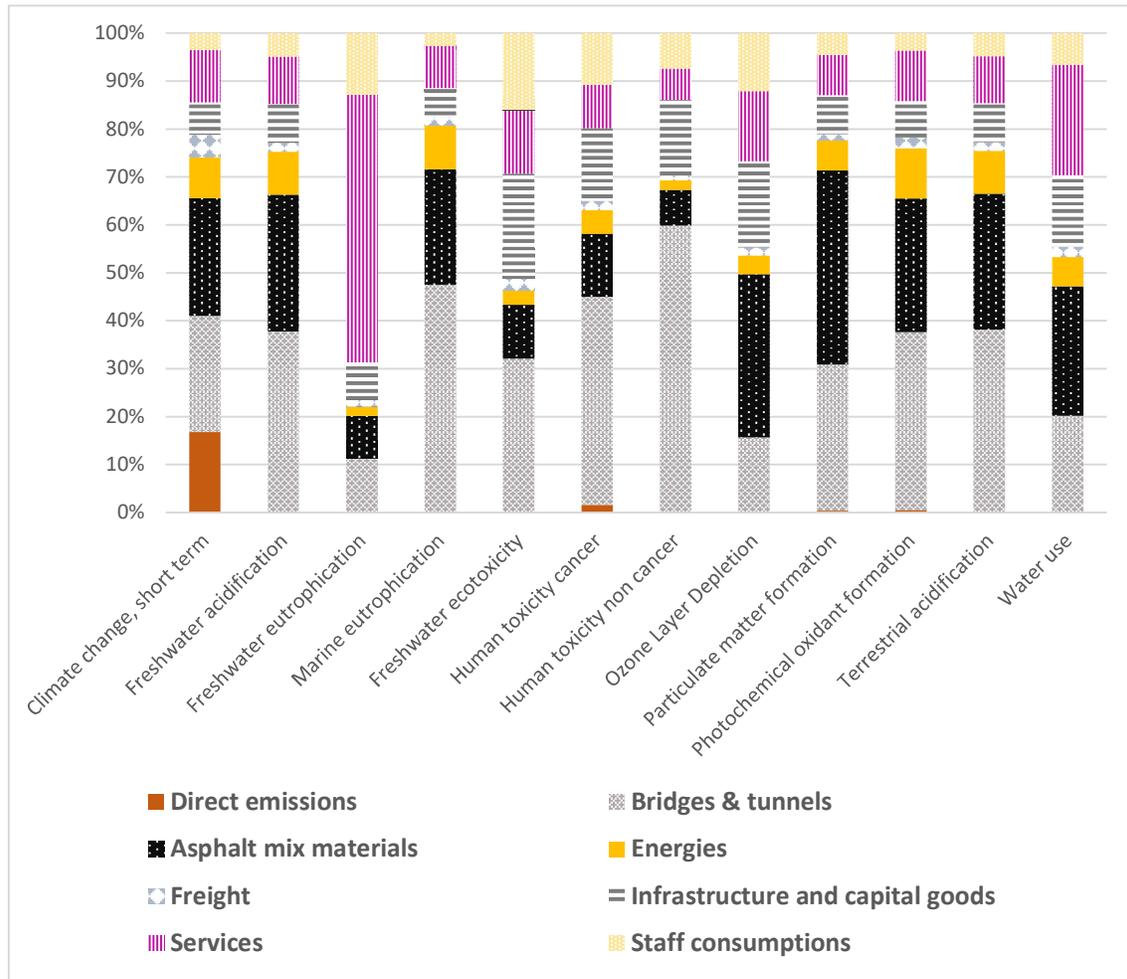

**Figure 4 Contribution of different activities to the different IW+ environmental impacts of the road industry at the midpoint level – overview**

**2.    Midpoint contribution details**

Figure 5 shows the contributions of direct emissions and purchases of the road industry to the different IW+ environmental impacts at the midpoint level. The figure illustrates the significant number of different factors to account for in the road industry when adopting a multicriteria approach. For instance, three different contributors need to be aggregated to represent two-thirds of the climate change impact (53%). These three contributors are the asphalt products (18%), the concrete (18%) and the direct emissions (17%). The other 20 sectors account for the rest of the impact (47%).

Asphalt (e.g. bitumen) and asphalt products as well as concrete are also major contributors to the impact of the road industry across multiple other impact categories: freshwater acidification (27 and 24%), carcinogenic human toxicity (8 and 21%), non-carcinogenic human toxicity (4 and 35%), marine eutrophication (17 and 39%), PM formation (19 and 20%), photochemical oxidant formation (21 and 29%), terrestrial acidification (23 and 28%) and water use (15 and 10%). While asphalt products account for binders and asphalt mixtures purchases (i.e. asphalt binder and aggregate production as well as asphalt mixing and products' transportation on the upstream life cycle of these products) related to pavement and sidewalks construction, concrete mainly connects to the construction of bridges, tunnels, and other structures (e.g. sidewalks and rare concrete pavements). We can conclude from Figures 4 and 5 that asphalt and asphalt products impacts must mainly be due to bitumen and asphalt mixing operations (e.g. burnt fuels).

Paper and paperboard purchases appear to be a negligeable contributor except on freshwater eutrophication (8%, Figure 5). While reducing paper consumption in companies is often promoted as an important environmental action, it appears insignificant for the road industry outside of freshwater eutrophication issues. Moreover, the crushing contribution of services on freshwater eutrophication (Figure 4) is brought by the purchase of external services, that accounts for 48% of the impact (Figure 5). Within the different external services bought, "Architectural, engineering and related services" and "Management, scientific and technical consulting services" contribute the most.

Steel purchases, machinery maintenance/repairs and other goods purchases (mainly wires and lighting fixtures) all significantly impact freshwater ecotoxicity and both toxicities (cancer and non-cancer).

Energy purchases by the road industry in Canada account around 10% of the contributions – summing natural gas, waste oil and other fuels. We note a negligible contribution from natural gas, while waste oil and other fuels – specifically heavy fuel oil – globally account for similar impacts. Waste oil does not appear particularly problematic on indicators accounting for the release of toxic substances such as human toxicities or ecotoxicity, while we know they constitute a serious risk (Government of Canada, 2018). This is probably because mostly small companies produce it, and that these small companies do not have to report to the NPRI. The direct emissions triggered by the combustion of fuels from production tools, machinery and company fleets, contribute specifically to climate change (17%) as specified previously.

Aggregates are particularly impacting on PM emissions (21%), despite the use of specific techniques to reduce dust, such as water spraying in some quarries especially close to densely populated areas.

Finally, concrete is an overall much higher contributor to the impacts than cement. It indicates that the industry mainly purchases concrete mixtures instead of manufacturing its own, as cement is the most carbon-intensive component of concrete. The cement bought is partly used in own-made concrete, and partly used to stabilize soils in pavement foundations.

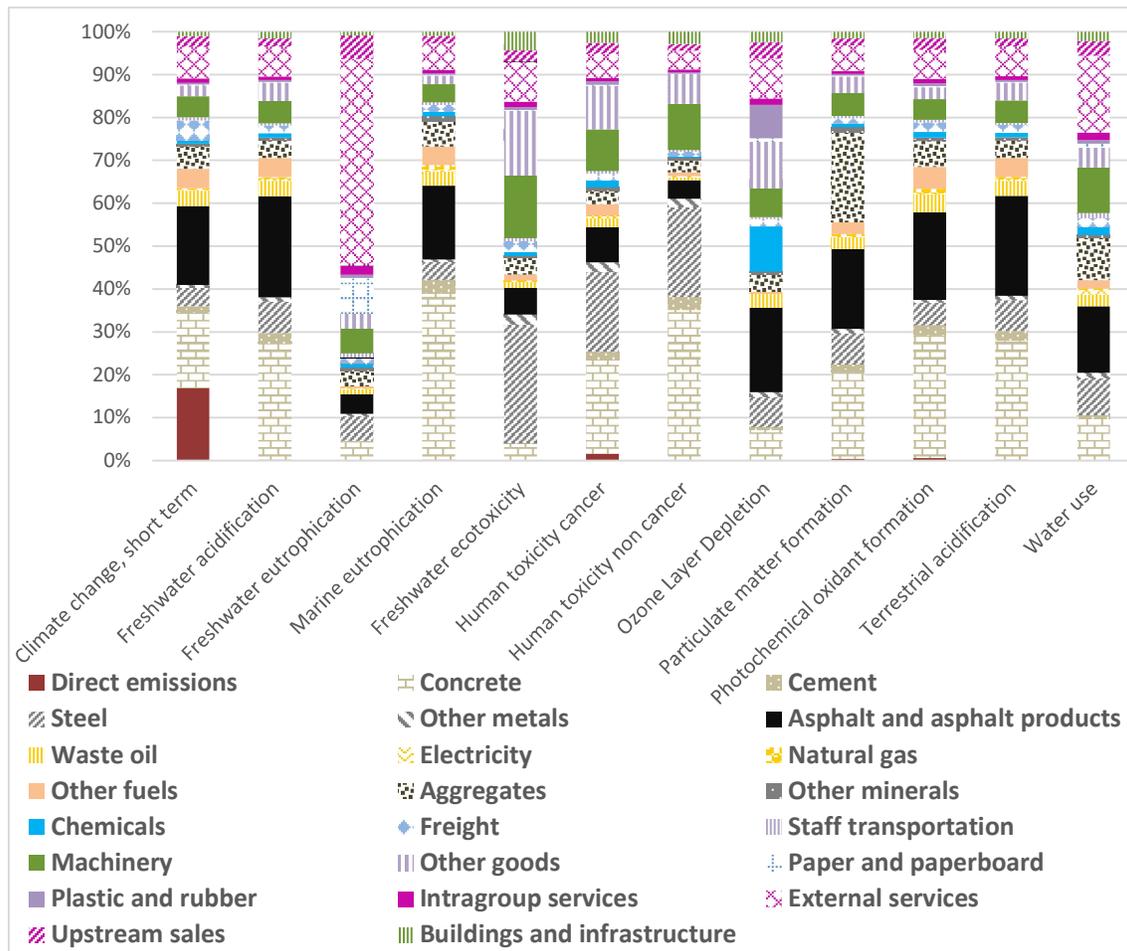

**Figure 5 Contributions of different activities to the different IW+ environmental impacts of the pavement industry at the midpoint level**

3. **Damage drivers**

Figure 6 shows the contribution of direct emissions and purchases by category of Canada's road industry for endpoint indicators. The main contributors are almost identical between the two kinds of damage. Contributors exceeding 5% are, by decreasing contribution: concrete (19% on EQ and HH), asphalt products purchases (18% on EQ and 17% on HH), direct emissions (17% on EQ and 16% on HH), as well as aggregates, freight, machinery, external services purchases, and steel (~5% each on each damage). The last third of the damage is due to the sixteen other kinds of purchases, with individual contributions equal to or lower than 4%. This analysis informs on which areas to focus the R&D topics and green purchases efforts. From a larger point of view, it also informs which LCIs to regionalize in priority in a road process-based LCA to enhance the quality of the results: the most important contributing processes must be regionalized first.

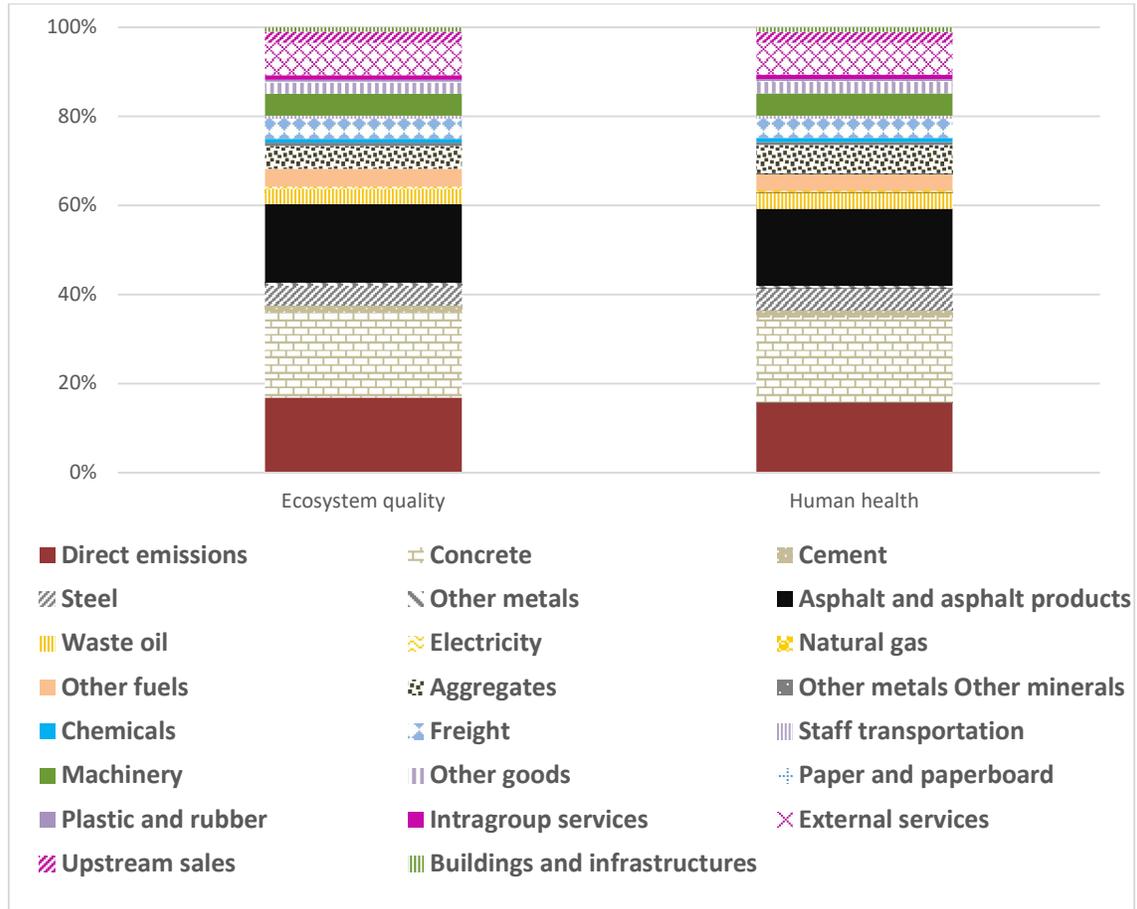

**Figure 6 Environmental drivers of Canada's road industry at the endpoint level**

### 4.2.3 *Metrics prioritization for green purchase plans*

Endpoint indicators in environmental quantification methods exist for decades (Steen and Ryding, 1992), and the pros and cons of midpoint and endpoint indicators in LCA have been largely debated, showing that both approaches have different adequate usages (Bare et al., 2000). One major benefit of the use of damage indicators in LCA is to offer a concise but complete overview of the environmental issues for non-LCA expert decision-makers. Nevertheless, damage indicators are rarely used, and an overwhelming number of studies only focus on a midpoint climate change impact, i.e. GHG emissions. Depending on cases, solely focusing on the climate change impact might still represent a decent estimate for endpoint categories, provided results are correlated. Figure 3 already showed that, at the sector scale, GHG flows represent the most impact at the endpoint level. Figure 7 now focuses on this correlation for each category of purchase by the road industry in Canada. A linear regression shows a very good correlation between the contribution of each category of purchase to the GWP and the two endpoint indicators in the case of the road industry, with a coefficient of determination $R^2$ equal to 0.995 and 0.991 resp. for ecosystem quality and human health. The figure shows that other fuels purchases (blue dot under the regression line in the ecosystem quality graph) have a particularly higher contribution to climate change than to the ecosystem quality endpoint indicator, while the opposite is true for aggregates and concrete (orange and brown dots above the regression line in the human health graph), indicating that these purchases have a bigger contribution to damages on human health than to climate change. Direct emissions impact is excluded from the analysis presented but its inclusion gives a higher $R^2$ (see SM). In the end, this analysis suggests that GWP is currently a good proxy to represent wider environmental impacts of Canada's road industry for each category of purchase and thus to support green purchase strategies for each category of suppliers. These conclusions must not be directly extrapolated to other sectors or locations, and they still depend on the openIO-Canada database and the IW+ characterization method. Finally, this example shows how to use EEIO analysis to assess the redundancy of indicators in an endpoint-oriented environmental strategy and reduce the number of metrics on which to support green purchases plans. This is a practical application concretely useful for industries and companies to actually implement green purchase plans, especially when sustainability teams do not have advanced LCA knowledge to conduct and interpret complex multicriteria assessments. Indeed, it reduces the complexity and the requests made to suppliers, while still allowing to select suppliers on a simple set of indicators. This would raise the greening strike force of consuming companies by a ripple effect on the entire upstream chain of suppliers.

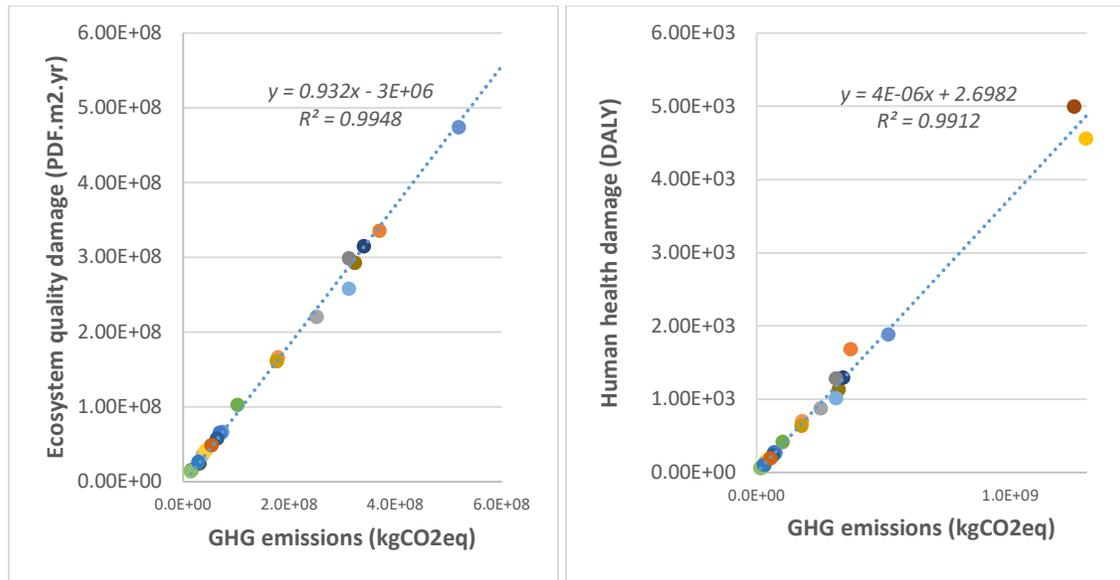

**Figure 7 Correlation between GWP (X-axis) and damages (Y-axis) per category of purchase from Canada's road industry**

## *4.3 Canadian roads' carbon footprint by life-cycle stage*

Previously, the breakdown by contributors of Canada's road industry was presented. In this section, the impact of the production/maintenance/end-of-life of the road industry (shown previously) is compared to the roads' usage by vehicles (downstream scope 3 according to the GHG protocol (WBCSD and WRI, 2004)), including vehicle tailpipe and fuel supply chain GHG emissions. The relative contributions of this comparison are presented in Figure 8, and absolute values are in Table 3. It shows that the construction, maintenance, and end-of-life of the Canadian road network only represent 5% of its life cycle carbon footprint, i.e. 8 Mt $CO_2$eq over the total 179 Mt $CO_2$eq from the Canadian road transportation system. Most of the carbon footprint (95%) comes from the pavement use stage - 80% of it from private vehicles (PVs) and 20% from freight services (see SM). PVs include private freight trucks as well as sport-utility vehicles (SUVs) and pickup trucks. Private freight trucks would account for one-third of the PVs emissions. SUVs and pickup trucks are very popular in Canada and account for around 60% of the household PV's direct emissions (Government of Canada, 2021c). Within the road network use stage, the vehicles' use stage accounts for 72% of the impact, while the manufacturing of vehicles represents 23% of it. Vehicles' maintenance shows negligible contribution. When looking at the contribution within the vehicles' lifecycle per category of vehicles, 85% of the PVs' life cycle emissions come from the vehicle use stage and 15% from the vehicle manufacturing amortization, while this stage accounts for 44% of the freight services' life cycle emissions (see SM). Within the vehicles' amortization emissions, PVs account for 50%, freight services vehicles for 38%, and light commercial vehicles (LCV) for 12%. Finally, if vehicles' use stage emissions are further broken down, we see that providing vehicle fuel generates 15% of the road transportation system GHG emissions, while PVs' tailpipe emissions account for 46% of it, and tailpipe emissions from freight services' vehicle 11%. On average, tailpipe emissions bring 79 % of the vehicle's use stage emissions, while the fuel supply chain brings the other 21% (see SM). This contribution is a bit higher than the 10 to 15% contribution from the fuel supply chain that can be found in databases such as the French ADEME's database (ADEME, 2019). This could be explained by the higher impact of the fuel on the Canadian market - partly coming from high carbon-intensive oil sands – compared to the impact of the European market's fuel (Charpentier et al., 2011; Masnadi et al., 2018).

To conclude, as the environmental carbon footprint of Canada's road network mainly comes from its use stage, the main lever of its decarbonization consists in a better management of the pavement-vehicle interactions (PVI) to reduce vehicle consumption and deterioration through

https://doi.org/10.1016/j.jclepro.2023.136039, accepted version, Journal of Cleaner Productionoptimally designed and maintained roads, rather than trying to reduce the impacts from the construction. This concretely means reducing the roughness of the road – for instance by controlling the popular International Roughness Index better –, as well as curves and slopes. This recommendation corroborates previous results at the road scale (de Bortoli, 2014; de Bortoli et al., 2022), showing that higher environmental impacts from a more intensive maintenance scheme can be more than offset by their consequential benefits on the use stage (de Bortoli et al., 2022), some previous studies focusing on the fuel consumption benefits rather than on the whole vehicle life cycle (Wang et al., 2014, 2012; Araujo et al., 2014). As these results show, the manufacturing of vehicles is already an important contributor to transportation impacts. It thus emphasizes the risk of burden-shifting due to the well-expected fleet electrification. Indeed, the current manufacturing of electric vehicles is more impactful – on GHG emissions as well as critical metal consumption - than those of combustion engine cars (ADEME et al., 2013; Roy and Ménard, 2016). This raises awareness about the necessity of carefully estimating the prospective impacts of current electrification policies for each region, to be sure that excess impacts from manufacturing are counterbalanced by higher benefits in the use stage. Finally, reducing the traffic in terms of kilometers traveled is obviously the most straightforward way to reduce the environmental impacts (of transportation), as in general reducing any consumption is.

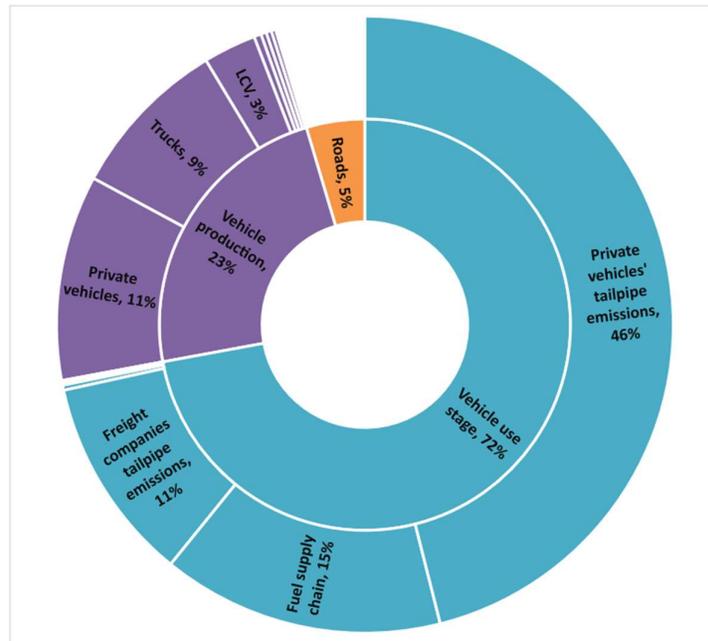

**Figure 8 GHG emissions drivers on Canada's road network**

**Table 3 GHG emissions from the Canadian road transportation system**

| Scope | Source | GWP (kg CO2eq) |
|---|---|---|
| Infrastructure | Roads | 8.23E+09 |
| Tailpipe emissions | Buses | 4.46E+07 |
| | LCVs | 4.73E+08 |
| | Freight services | 1.92E+10 |
| | Tourism | 2.94E+07 |
| | School buses | 1.58E+08 |
| | Taxis | 2.19E+08 |
| | Private vehicles | 8.24E+10 |
| Fuel supply chain | Fuel | 2.64E+10 |
| Manufacturing | Buses | 1.58E+08 |
| | Trailers | 6.75E+08 |
| | LCVs | 4.96E+09 |
| | Companies' trucks | 1.52E+10 |
| | Tourism vans | 5.09E+08 |
| | Other vehicles | 4.13E+08 |
| | Private vehicles | 1.93E+10 |
| | Maintenance | 4.91E+08 |
| TOTAL | **TOTAL EMISSIONS** | 1.79E+11 |

# 5 Discussion

## 5.1 The environmental transition of the Canadian road industry

### 5.1.1 Insights for the road industry actors

Our results highlight how EEIO analysis can help quickly sketch an environmental transition plan by first understanding the multicriteria environmental responsibility of an industry at the national scale, also accounting for embodied emissions in the supply chain – including imports, revealing the main contributors to these impacts, and reducing the environmental dimensions to consider in R&D and green purchase plans. This case study shows the major roles of purchases of concrete and asphalt products, as well as direct emissions on the climate change midpoint indicator for Canada's road industry (18, 18 and 17% of the impact). As regards direct emissions, and since the climate change midpoint indicator also brings a major contribution at the endpoint level, it calls for efforts on the energy consumed by production tools, machinery, and company fleets. As such, switching from heavy fuel to natural gas would for instance reduce these emissions by 31% per megajoule consumed due to lower carbon factors (Quebec Ministry for the Energetic transition, 2019). Biofuels also generally show lower impacts than fossil fuel but can generate burden-shifting (Jeswani et al., 2020), calling for studying carefully their local and global impacts from specific sourcing. Electrification of construction machinery, company fleets, and production processes is a much more promising avenue where electricity mixes are low-carbon such as in Quebec, yet subject to an increase in electric capacity respecting electricity decarbonization. Our results also emphasize the importance of engineering structures on the overall impact of infrastructure which corroborates railway process-based LCAs (de Bortoli et al., 2020; Chang and Kendall, 2011). This contribution would mainly be due to the GHG emissions of concrete (and more specifically of cement) and metals, building machines having a limited contribution to this impact (de Bortoli et al., 2020). Moreover, this limited impact of the construction process compared to the provision of materials has also been demonstrated in many road LCAs (e.g. the meta-analysis by de Bortoli, 2020). More generally, the contributions of the different types of purchases made by the road industry in Canada will enable this industry to prioritize the development of a green purchasing charter for categories with a strong environmental impact, which concerns - apart from engineering structures – asphalt mixture and materials to produce them. When looking further at the environmental impact from the "Asphalt (except natural) and asphalt products", EEIO results show that 43% of the carbon footprint comes from purchases in asphalt binder, 22% from fuel purchases, 10% from aggregates, 9% from crude oil purchases (aiming at producing asphalt binders and fuels), and 5% from freight. This embodied carbon structure is consistent with previous studies (de Bortoli, 2020) and confirms the prevalence of asphalt binders and fuels for asphalt mixing and asphalt binder production in the emissions from asphalt mixture production.

Finally, we used EEIO analysis at the national industrial scale, but it can also be a powerful tool at the company scale to hierarchize environmental actions. Indeed, EEIO analysis can also be statically used internally by companies to screen environmental levers based on their buying reports, or even dynamically to simulate the environmental impacts of a change in the purchases of the company, relating to a change in the buying choice, the materials or services sold, or the production process. This process must involve all the departments of a company: R&D, material, buying, and production departments.

*5.1.2   The trifling contribution of road construction compared to road usage*

Quantifying the environmental contribution of each life-cycle stage of the road to its lifetime impact is a recurring question in the literature, especially to understand to which extent the pavement use stage is important, as it is often excluded from road LCAs scope (Santero et al., 2011). Several studies emphasized that the vehicle pipe emissions carry a crushing contribution to the road life-cycle impact (de Bortoli, 2018; Wang et al., 2014, 2012; Chappat and Bilal, 2003). When included, the impact of vehicle manufacturing and maintenance has shown to be also important in the rare studies accounting for it, for instance representing 21% of the primary energy consumed by a road system over 30 years (de Bortoli, 2014) or around 15% of the carbon footprint of passenger car modes (de Bortoli and Christoforou, 2020; Chester and Horvath, 2009). However, these assessments were always carried out on a particular road or a particular mode (e.g. passenger car transportation), and a more global overview at the network level was missing. The environmental picture that we give for Canadian roads demonstrates that decarbonizing the road construction industry might not be a priority to tackle the impact of road transportation, contrary to building and maintaining roads to lower vehicle consumption and tear and wear, or directly reducing vehicle manufacturing and use's GHG emissions. Results show that fleet replacement is far from being a secondary question in terms of road's climate transition, which may be explained by the fact that cars in Canada are the second heaviest in the world, weighing 1717 kg on average in 2017 (IEA, 2019). Fleet replacement is a particularly thorny question with the penetration of electromobility, as electric vehicles emit more at the manufacturing stage but less during the use stage than internal combustion engine (ICE) vehicles, which can lead to lowering GHG emissions on the entire life cycle as shown in Quebec (Roy and Ménard, 2016).

### 5.2  Limitations and future work on the openIO-Canada model

*5.2.1  Overview*

OpenIO-Canada has been designed with a few inherent limitations and assumptions that are made clear in this section. First, the physical flow accounts of Statistics Canada (used to quantify GHG emissions and water use) are provided at a more aggregated level (240 sectors) than the economic data that is used by openIO-Canada (492 sectors). Hence, an economic allocation approach was adopted to distribute GHG emissions and water use. For instance, physical flow accounts only provide GHG emissions for the "Crop and animal production" sector. OpenIO-Canada thus considers any more refined subsector that belongs to the broad "Crop and animal production" sector and uses the weighted average of their sales share to determine their emission level. In other words, if the "Wheat" sector represents 6% of the total sales of "Crop and animal production", the "Wheat" sector will be attributed 6% of the total GHG emissions of "Crop and animal production". This approach triggers inconsistencies as the emissions do not follow sales, for example, in the EXIOBASE database, the GHG emission factor for the wheat sector is 1.56kg$CO_2$eq/€ while the one for cattle is 4.54kg$CO_2$eq/€.

Second, openIO-Canada does not provide a version with capital goods endogenized. Capital endogenization is a method used in EEIO analysis to distribute the impacts of capital formation (e.g., the construction of a machine) to the different sectors of the economy requiring the formation of capital. Using an example to make it clearer, openIO-Canada records the construction of a building (and its associated emissions) but does not link the construction of the building to the economic sector that required the building. Therefore, contribution analyses on sectors provided by openIO-Canada do not include the formation of capitals, but rather only include recurring operations such as maintenance and repairs.

Third, international imports are considered produced as in the importing province, as usually done in EEIO databases. As an example, an American car imported in Quebec will thus be considered produced the same way as a car directly produced in Quebec, both for economic and environmental flows. This approach is called "DTA" for "Domestic Technology Assumption", and is a issue specific to single region input-output approaches. Integrating the openIO-Canada table into a global multi-regional input-output (MRIO) system would allow for regionalizing import impacts.

Fourth, the NPRI has some limitations. This readily available Canadian national emission and consumption database does not cover all the flows accounted for in LCA databases, such as ecoinvent. While NPRI covers a fair amount of 323 types of substances, some flows are nevertheless missing, such as land occupation/transformation, ionizing radiations, water consumption, as well as mineral and fossil resource use. Moreover, the NPRI covers emissions of the 8000 most important complying industrial sites across Canada. Pollutant emissions (other than GHG emissions) are thus underestimated, as many smaller industrial sites are not covered by the NPRI.

Fifth, EEIO analysis is a powerful tool to quickly assess large-scale systems but is rather limited in its ability to assess disruptions such as technical or material innovations or new national production sectors. A few studies manually simulated new productions, such as Leurent and Windisch (2015) did in an economic input-output analysis to assess the macroeconomic impacts of a potential French electric battery production sector, while other coupled EEIO databases to equilibrium models to simulate disruptions van Sluisveld et al. (2016). Lately, the industrial ecology community rather turned toward prospective LCA, a growing corpus of studies being dedicated to these assessments (Thonemann et al., 2020).

### 5.2.2 *Focus on the underestimated non-GHG direct emissions*

As specified in the limitations of openIO-Canada, the NPRI only covers a part of Canada's industrial sites. Indeed, according to the Canadian Environmental Protection Act, only the "owners or operators of facilities that meet published reporting requirements are required to report to the NPRI" (Government of Canada, 2020a). These requirements are a facility with more than 10 full-time employees - i.e. 20 000 hours of work per year – and carrying out specific activities, including quarry operations and operation of stationary combustion equipment (Government of Canada, 2021d). In Canada, around 8000 facilities report their pollutant releases in the 2017 NPRI, based on reporting requirements that mainly consider the type of activities, the total number of hours worked and substances manufactured (Government of Canada, 2011). But, in 2020 for instance, the manufacturing sector accounted for 90359 establishments, and the oil and gas extraction sector only 4125 (Government of Canada, 2021a). Thus, the completeness of the activities covered by the NPRI database is rather poor in number of sites but may be good in the percentage of production covered as it includes major producers. Nevertheless, we do not have access to this information yet, and this flaw in the coverage leads to an unknown underestimation of the substances assessed with the NPRI (i.e., all substances except GHGs, the latter being assessed with physical accounts). As a result, the estimation of the climate change indicator reliably covers the whole Canadian economy while the estimation for other impact categories may not. This, to date, limits somehow the robustness of some conclusions, for instance the correlation between climate change and endpoint categories shown in section 4.2.3. Indeed, if only climate change is characterized comprehensively, it creates a bias towards its contribution in both endpoint categories. Moreover, the overwhelming contribution of climate change to the damage to ecosystems and human health (section 4.1.2.) could result from the truncation of the NPRI data used in the current version of openIO-Canada. It could also explain the low contributions of direct emissions to midpoint indicators (section 4.2.2.). Assessing this bias will require improving the coverage of non-GHG emissions by openIO-Canada in the future, nonetheless, all the methodological aspects proposed to use EEIO analysis to sketch environmental transition plans are totally valid.

## 5.3  Articulation between EEIO analysis and LCA for transition plans

### 5.3.1  EEIO analysis as a screening tool to build upon for LCA

EEIO analysis can be a screening tool for transition plan development, but it can also be used to orientate LCA methodological choices such as system boundaries or LCI regionalization, by looking at the contribution of specific products and services calculated with EEIO databases, as detailed below.

What is usually considered outside the system boundaries of road process-based LCAs – namely staff consumptions, services, and sometimes infrastructure and capital goods amortization – accounts according to openIO-Canada for 15 to 73% of the impacts, resp. on marine eutrophication and freshwater eutrophication. In particular, external services are always excluded from these studies and bring between 4 and 48% of the impacts, resp. on human non carcinogenic toxicity and freshwater eutrophication. EEIO analysis is thus an interesting tool to rapidly screen the impact of time-consuming aspects to model in process-based LCAs and decide whether to include them in the system boundaries of an LCA in a more specific study. Amortization of infrastructure, machinery, and buildings is more often included, especially in the ecoinvent database. Yet the models are often very generic and/or outdated, as illustrated by the quarries: a single model for each type of site (loose rocks or solid rocks) was created between the end of the 90s and the beginning of the 2000s, based on 4 Swiss quarries (Kellenberger et al., 2007). However, the impact of the depreciation of infrastructure is even more important as the carbon intensity of energies decreases. If the contribution of this depreciation is already significant, it will undoubtedly be more and more so, which calls for improving the classic LCA models. Note, all the same, that our EEIO database on this point has limits: we consider the impacts of the year 2017 to be representative, while there can be strong heterogeneities depending on the year of purchase. Capital goods should also be endogenized in openIO-Canada as it has been done in EXIOBASE (Södersten et al., 2018) and US-EEIO (Miller et al., 2019).

It may also be interesting to use the major contributors to the environmental impacts highlighted through EEIO analysis to prioritize the regionalization of life cycle inventories (LCIs) for a sector, particularly if the national market differs from the markets modeled in existing databases, or if national technologies do not coincide with existing models in the case of domestic production.

However, as EEIO analysis is in most cases entirely based on economic allocations (except for a few hybrid-unit versions), the importance of the contributors found with EEIO analysis could be distorted compared to LCA contributors, as LCA results are more disaggregated than EEIO results, and mainly uses non-economic allocation methods such as mass or other physical allocations to deal with multifunctional systems. Thus, comparing EEIO analysis and process-based LCA results, and using both types of methodologies to interpret and use results in decision-making support could bring more robustness on the environmental structure and levers of a system, or more insights on the uncertainties to deal with. EEIO analysis and LCA must then be used complementarily to strengthen environmental decision processes.

*5.3.2  Uncertainty and data quality in EEIO analysis versus LCA*

EEIO databases generally cover a more limited number of substances than LCI databases which typically consider more than a thousand substances. Nevertheless, there is great variability in the magnitude of the environmental consequences of substances. For instance, with only about 35 pollutants covered, EXIOBASE has been found to still reliably cover many impact categories apart from those related to toxicity (Muller, 2019).

Another potential drawback of EEIO analysis pertains to the reliability of national pollutant inventories, used by EEIO databases and compiled by government agencies, which rarely stem from actual measurements on industrial sites. In the case of Canada for example, certain specific emissions must be measured on an annual or multi-year basis, e.g., the emissions of certain pollutants for facilities burning used oil in Quebec. Most emissions on the other hand are simply reported based on calculation methodologies prescribed by the Government of Canada which may be incomplete or obsolete. As an illustration, a calculator is made available for asphalt plants to estimate their pollutant emissions (Government of Canada, 2020b). Parameters - such as annual production, fuel consumption, type of production, or asphalt mixing temperature to name but a few - are fairly comprehensive and make it possible to take into account technologies and specific equipment. However, according to this "hot mix asphalt plant" calculator's documentation, its emission factors are based on US EPA data published between 1994 and 2005. Moreover, the calculator does not differentiate emissions from a boiler using waste oil and oil #6, while used oil facilities need to be tested regularly due to the risk related to specific metals and contaminants emissions. It, therefore, seems necessary to increase the consistency of emission reports, for example by providing updated calculators. The more so as a specific calculator also exists for waste oil combustion (Government of Canada, 2018), but the existence of this calculator is not pointed out in the section relating to the production of asphalt mixtures. Finally, efforts should be put on the quality of these databases, in terms of completeness, reliability, as well as technological, geographical, and temporal representativeness. Moreover, giving access to the information on the quality of the background data could allow analysts to assess the robustness of their assessments. This data quality problem also concerns LCA but is commonly addressed in LCA through data quality characterization and various uncertainty propagation methods (Baker and Lepech, 2009), while it is more rarely the case in EEIO analysis (see the example by Lenzen et al. in 2010).

# 6   Conclusions

EEIO analysis must be used as one tool to orientate the environmental transition plans of industries as proved by the example of the pavement industry in Canada. First, it estimates faster than LCA the multicriteria environmental contribution of an industry for cradle-to-gate on a large scale such as a country: in our illustration, the road industry accounts for around 1.0% of most of the country's damages on a consumption-based approach, i.e. 10% of the construction sector impacts. It thus unearths industrial key drivers on which organizing the environmental transition through green purchase and R&D prioritization strategies. The road industry in Canada must reduce (1) its direct emissions through the investment in new machinery using cleaner energies such as low-carbon electricity, biomass, and natural gas, and (2) the impact of material purchases, especially concrete and asphalt products. Second, EEIO analysis spots the critical midpoint indicators explaining most of the damage to areas of protection such as ecosystems and human health, allowing for reducing the set of classical LCA indicators to monitor ecological transition plans, multicriteria assessments being less understandable for non-experts as the number of metrics rises. Climate change dominates the midpoint category contributions to damages in the pavement sector, followed by marine acidification, and PM formation due to aggregates. Third, EEIO analysis rapidly seizes often neglected sources of impacts in traditional LCA: capital goods amortization, staff consumption, and services. It must be used to set the system boundaries adapted to specific LCAs. Finally, it helps estimate the impacts of some types of goods at a large scale on a cradle-to-grave perimeter: its application to Canada's road network confirms that the use stage of the roads is capital to its environmental impact (95% of the GHG emissions), mainly due to vehicle tailpipe emissions (72% of the total impact) but also to vehicle amortization (23%). The construction and maintenance of the roads contribute little over the life cycle (5%), and the major lever for decarbonization is expected from better managing PVI to reduce the use stage impacts through better designed and maintained roads.

**Acknowledgement:** the authors want to thank Ivan Drouadaine, Amelie Griggio, and Marc Proteau for funding this project and supporting it technically. They also thank Jimmy Mikedis, Supervisor of the Data Service Centre of Statistics Canada, Anna Hatzihristidis, Consulting analyst at the Statistical Information Service of Statistics Canada, as well as their colleagues, for their help in navigating Canada's use and supply table classification codes. They are also grateful to Bitume Quebec, and especially Stéphane Trudeau, technical director, for exchanging on the market of bitumen and roads in Canada.

**Funding source and role:** this study has been conducted under the research project HEATI (Harmonized Environmental Assessment of Transportation Infrastructure) funded by the technical department of Eurovia – VINCI group. The work has been hosted at the Americas Technical Center of Eurovia Canada.

**Data statement**: OpenIO-Canada is accessible online: https://github.com/CIRAIG/OpenIO-Canada. The case study calculation algorithm is available on Zenodo (https://doi.org/10.5281/zenodo.6505413). Excel spreadsheets of category aggregation, as well as raw and worked results, are also made available in the supplementary material.

**Authors' contribution**: Conceptualization: ADB; Methodology: ADB, MA; Data curation: ADB, MA; Formal analysis: ADB; Software: MA; Validation: ADB; Visualization: ADB,

MA; Roles/Writing - original draft: ADB; Writing - review & editing: MA; Funding

acquisition: ADB.